\documentstyle[seceq,epsf,wrapfig,twoside]{ptptex}
\notypesetlogo  
\title{Weak Radiative Decay $\Lambda_b \to \Lambda ~\gamma $ 
and Quark-Confined Effects\\ 
in the Covariant Oscillator Quark Model }
\author{%
Rukmani {\sc Mohanta}, Anjan K. {\sc Giri}, Mohinder P. {\sc Khanna},\\
Muneyuki {\sc Ishida}$^{*}$ and Shin {\sc Ishida}$^{**}$
}
\inst{%
Department of Physics,
Panjab University, Chandigarh-160014, India\\
$^{*}$Department of Physics, Tokyo Institute of Technology\\
Tokyo 152-8551, Japan\\
$^{**}$Atomic Energy Research Institute, 
College of Science and Technology\\
Nihon University, Tokyo 101-0062, Japan\\ 
}
\recdate{%
May 30, 1999
}
\abst{%
Motivated by the observation of the decay $B \to K^* ~\gamma $
by the CLEO collaboration, we have systematically analyzed the weak 
radiative decay $\Lambda_b \to \Lambda ~ \gamma $,
evaluating the confined effects of quarks in the 
covariant oscillator quark model.
This decay process receives both short distance
(electromagnetic penguins at the one loop level) the long distance 
contributions in the quark level. 
The long distance contributions are analyzed using the
vector meson dominance (VMD) method. The estimated branching ratio
is found to be $0.23 \times 10^{-5}$.
}

\begin{document}
\maketitle

\setcounter{tocdepth}{4}

\section{Introduction}

Weak radiative (flavor changing neutral current) decays of hyperons
have attracted the interest of physicists during the last three
decades. In the Standard Model (SM) these processes are forbidden 
at the tree level and are strongly suppressed by the GIM mechanism.
Hence they offer a unique possibility to test the CKM sector of the
SM and possibly open a door to physics beyond it. Experimental
data are now available for the light baryon sector, i.e., $\Sigma^+
\to p~\gamma $, $\Lambda \to n ~\gamma $, $\Xi^- \to \Sigma^- ~ \gamma$,
$\Xi^0 \to \Sigma^0 ~\gamma $ and 
$\Xi ^0 \to \Lambda ~\gamma $,\cite{ref1}
which involve transitions of the type $ s \to d \gamma $.
Recently the rare decay $B \to K^* ~\gamma $ has been observed by the CLEO
collaboration,\cite{ref2} which is dominated by the quark level 
process $ b \to s \gamma $. Therefore, one may expect that there
is also some possibility for the rare decay of heavy baryons.
For the heavy hadron decay processes of this type, it is considered
that the confined effects of quarks play generally an important
role, since there is a large difference between the initial and 
final hadron masses. In a preceding paper\cite{ref31} we  
investigated the rare $B\to K^*\gamma$ process, evaluating the confined 
effects in the framework of the covariant oscillator quark model
(COQM).\cite{ref4}

In this paper we study
the weak radiative decay $\Lambda_b \to \Lambda ~
\gamma $, which results from the interplay of electroweak and gluonic
interactions. At the quark level, there are two essential mechanisms
responsible for the weak radiative decays $(b \to s \gamma)$
: the short distance (SD)
electroweak penguin and the long distance (LD) contributions.
Recently, an investigation\cite{ref3} of long distance
contributions to these decays was made 
using vector meson dominance (VMD)
at the quark level as $ b \to s [\psi ]$ followed by the conversion
$[\psi ]\to \gamma$. We estimate the contributions
to the decay  $\Lambda_b \to \Lambda ~\gamma $ arising
from both sources, evaluating  the hadronic matrix elements
in COQM.\cite{ref4}

One of the most important motivations for the COQM is to covariantly 
describe the center of mass motion of hadrons, preserving the
considerable successes of the nonrelativistic quark model on the
static properties of hadrons. A key in COQM for doing this
is to treat the square masses of hadrons as opposed to the mass
itself, as done in conventional approaches. This makes the covariant
treatment simple. 
The COQM has been applied to various problems
\cite{ref12b} with satisfactory results. Recently, Ishida et al.
\cite{ref13b,ref181b} have studied the weak decays of heavy 
hadrons using this model
and derived the same relations of weak form factors 
for heavy-to-heavy transitions as in HQET.\cite{ref2b} 
In addition, the COQM is also applicable to heavy-to-light
transitions. As a consequence, this model does incorporate
the features of heavy quark symmetry and can be used to
compute the form factors for heavy-to-light transitions as well,
which is beyond the scope of HQET. Actually, in previous
papers we made analyses of the spectra of exclusive 
semi-leptonic\cite{ref181b} decays of $B$-mesons,
of non-leptonic decays of $B$ mesons,\cite{meson} and 
of hadronic weak decays of 
$\Lambda_b$ baryons\cite{baryon} along this line of
reasoning, leading to encouraging results.

The paper is organised as follows. In \S 2 we present the 
methodology necessary for our analysis. The short distance and
long distance effects are discussed in \S 
 2.1 and 2.2 and evaluation of the hadronic matrix elements are given in 
\S 2.3. Section 3  contains our results and discussion.

\section{Methodology}

The general amplitude of baryon weak radiative decay is
given by
\begin{equation}
{\cal M}(B_i \to B_f~\gamma)=i \bar u_f (a+b \gamma_5)
\sigma_{\mu \nu} \epsilon^{\mu} k^{\nu} u_i\label{eq:as1}
\end{equation}
where $u_i\ (u_f)$ is the Dirac spinor of the initial (final) baryon,
$\epsilon^\mu\ (k^\nu )$ denotes the polarization (momentum) vector
of the photon, and 
$a$ and $b$ are parity-conserving and parity-violating
amplitudes, respectively. The corresponding decay rate is given as
\begin{equation}
\Gamma (B_i \to B_f~ \gamma) =
\frac{1}{8 \pi} \left (\frac{M_i^2 -M_f^2}{M_i}
\right )^3 (|a|^2 +|b|^2)\;,
\end{equation}
and the asymmetry parameter is given as
\begin{equation}
\alpha = \frac{2 {\rm Re}~( a^* b)}{(|a|^2 +|b|^2)}
\end{equation}

\subsection{Short distance contribution}

The effective Hamiltonian \cite{ref6} for the short distance 
$b \to s $ transition including the QCD correction is given by

\begin{equation}
{\cal H}_{\rm eff}^{SD}(b \to s \gamma) = -\frac{G_F}{\sqrt{2}}
\frac{ e}{16 \pi^2}  F_2 V_{tb}V_{ts}^* F_{\mu \nu}~ \left [
m_b \bar s ~\sigma^{\mu \nu}~(1+\gamma_5)~b
+m_s \bar s ~\sigma^{\mu \nu}~(1-\gamma_5)~b \right ]
\;,
\end{equation}
where $F_{\mu \nu} $ is the electromagnetic field strength tensor,
$V_{ij} $ are the CKM matrix elements, $F_2 \approx
F_2(x_t) -F_2(x_c) \approx F_2(x_t) $ with $x_i=m_i^2/M_W^2 $ and
\begin{equation}
F_2(x)=\rho^{-16/23} \left \{ \bar F_2(x)+\frac{116}{27}\left [
\frac{1}{5}(\rho^{10/23}-1)+\frac{1}{14}(\rho^{28/23}-1)
\right ] \right \}
\end{equation}
with
\begin{equation}
\bar F_2(x)=\frac{(8x^2+5x-7)x}{12(x-1)^3}-\frac{(3x-2)x^2}{2(x-1)^4}
{\rm ln}x \;,
\end{equation}
and
\begin{equation}
\rho=\frac{\alpha_s(m_b^2)}{\alpha_s(M_W^2)}=1+\frac{23}{12 \pi}
\alpha_s(m_b^2)~
{\ln}\left (\frac{M_W^2}{m_b^2} \right )\;.
\end{equation}

Numerically $F_2(x_t)=0.65 $ for $\Lambda_{\rm QCD}=200$ MeV and
$m_t=174$ GeV. Furthermore, using the unitarity of the CKM matrix,
i.e., $V_{tb} V_{ts}^* = -V_{cs}^* V_{cb}-V_{us}^* V_{ub} $ since
$V_{us}^* V_{ub} << V_{cs}^* V_{cb} $, we obtain the matrix element 
arising from the short distance effects as

\begin{equation}
\langle \Lambda \gamma | {\cal H}_{\rm eff}^{SD}|\Lambda_b \rangle
=i \frac{G_F}{\sqrt 2} \frac{e}{8 \pi^2} F_2 V_{cb} V_{cs}^* \epsilon_\mu
k_{\nu} \langle \Lambda | \bar s [ m_b \sigma^{\mu \nu} (1+\gamma_5)
+m_s \sigma^{\mu \nu} (1-\gamma_5)]b |\Lambda_b \rangle \;.
\label{eq:eqn12}
\end{equation}

\subsection{Long distance contribution}

The long distance contributions were recently estimated using the
vector meson dominance method in Ref.~\citen{ref3}. 
At the quark level, this assumes the dominance
of the process
 $ b \to s [\sum_i\psi_i]  \to s ~ \gamma $, where 
all the $\bar c c$, $J=1$ excited as well as ground  charmonium states
are taken into acount as $\psi_i $. The relevant part of the
effective Hamiltonian describing the process is given by
\begin{eqnarray}
&&{\cal H}_{\rm eff} = \frac{G_F}{\sqrt{2}}\; V_{cb}\;V_{cs}^*\;
[C_1(\mu)~ O_1 + C_2(\mu)~  O_2 ], \label{eq:1}
\end{eqnarray}
with
\begin{eqnarray}
&& O_1=(\bar s c)^{\mu}\; (\bar c b)_{\mu}~~~~~~~~~~
\mbox{and}~~~~~~~~~~~ O_2=
(\bar s b)^{\mu} (\bar c c)_{\mu}\;,
\end{eqnarray}
where the quark current $(\bar q_i q_j )_{\mu}=
\bar q_i \gamma_\mu(1-\gamma_5) q_j$ denotes
the usual $(V-A)$ current. 

Using factorization, we obtain the inclusive decay amplitude
for the process $ b \to s \psi $ as
\begin{equation}
{\cal M}(b \to s \psi (k_1, \epsilon_1))= \frac{G_F}{\sqrt 2}
V_{cs}^* V_{cb} a_2(\mu) f_{\psi}(m_{\psi}^2) m_{\psi }
\bar s \gamma^{\mu}(1 -\gamma_5) b \epsilon_{1 \mu}\;,
\end{equation}
where $a_2(\mu)=C_2(\mu)+C_1(\mu)/N_c $, and $N_c  =3$ is the
number of colors. $k_1$ and $\epsilon_1 $ are the momentum and 
polarization vector of the vector meson $\psi $. In the above
equation we have used the matrix element
\begin{equation}
\langle \psi(k_1,\epsilon_1) |(\bar c c)^{\mu} |0 \rangle
= f_{\psi}(m_{\psi}^2)~m_{\psi} \epsilon_1^\mu .
\end{equation}
We have used the value $a_2^{\rm eff}$, which is determined experimentally
from the world average branching ratio of $\bar B \to K^* \psi $
\cite{cheng1} as
\begin{equation}
a_2^{\rm eff}=0.23\;.
\end{equation}

Now we wish to replace the $\psi $ meson with the photon and
construct a gauge invariant amplitude. This can be done by
eliminating the longitudinal component of the $\psi $ meson so that
$\epsilon_1^{\mu} $ is changed to the polarization vector of the
photon $\epsilon^\mu $. For this purpose we use the procedure
of Golowich and Pakvasa.\cite{pakvasa1} 
Now, using the equation of motion for the $b$ 
quark, i.e. $\not\! p b =m_b b $, and momentum conservation, 
$p=p^{\prime}+k_1 $, we obtain
\begin{equation}
\bar s \gamma_\mu (1-\gamma_5)b = \frac{1}{m_b}[\bar s \gamma_{\mu}
\not\!{p}^\prime (1+\gamma_5) b+\bar s \gamma_{\mu}\not\!{k}_1
(1+\gamma_5) b],\label{eq:eqn7}
\end{equation}
where $p$ and $p^\prime $ are the momenta of the $b$ and $s$ quarks
respectively. The contribution due to the first 
term in Eq. (\ref{eq:eqn7}) is neglected, since $m_s \ll m_b $ and
$p^{\prime \mu} \epsilon_{1 \mu}^T=0 $, which follows
from $\epsilon_{1 \mu}^T p^{\mu}=0 $ in the rest frame of
the $b$ quark and the transversality condition $\epsilon_{1 \mu}^T 
k_1^\mu $=0, where $\epsilon_{1 \mu}^T $ is the transverse
polarization vector of the $\psi $ meson. The second
term can be written as
\begin{equation}
\frac{1}{m_b} \bar s \gamma_\mu \not\!{k}_1
(1+\gamma_5)b = \frac{1}{m_b}\{\bar s (1+\gamma_5)k_{1\mu}b
-i \bar s \sigma_{\mu \nu}k_1^\nu (1+\gamma_5)
b \}\;.\label{eq:mat1}
\end{equation}
In Eq.~(\ref{eq:mat1}) only the $\sigma_{\mu \nu } $ term couples
to the transverse component of $\psi $, and we obtain the corresponding
amplitude as 
\begin{equation}
{\cal M}( b \to s \psi^T)=-\frac{G_F}{\sqrt 2}V_{cb}V_{cs}^* a_2(\mu)
f_{\psi}(m_{\psi}^2) \frac{m_{\psi}}{m_b} \bar s \sigma_{\mu \nu}
(1+\gamma_5)b \epsilon_1^{T \mu} k_1^\nu \;.
\end{equation}
For the $\psi^T \to \gamma $ conversion following the VMD mechanism,
we have
\begin{equation}
\langle 0| J_\mu |\psi^T(k_1, \epsilon_1^T) \rangle =e Q_c
f_{\psi}(0) m_{\psi} \epsilon_{1 \mu}^T,
\end{equation}
where $Q_c=2/3 $ and $f_{\psi}(0) $ is the coupling at $k_1^2=0$.
Using the intermediate propagator of the $\psi $ meson at $k_1^2=0$,
we get
\begin{equation}
{\cal M}(b \to s \psi^T \to s \gamma )
=i\frac{G_F}{\sqrt 2} ~V_{cb} V_{cs}^* ~a_2 ~f_{\psi}^2 (0)~\frac{e Q_c}
{m_b}~\bar s ~\sigma_{\mu \nu}~(1+\gamma_5) ~b ~
\epsilon^\mu ~k^\nu \;.\label{eq:eqn8}
\end{equation}
It should be noted that the coupling structure is the same as that due to
the short distance electromagnetic penguin operator.
The expression for the amplitude Eq. 
(\ref{eq:eqn8}) can be completed by summing over all the $c \bar c $ 
resonant states $\psi(1S)$, $\psi(2S)$, $\psi(3770)$, $\psi(4040)$, 
$\psi(4160)$ and  $\psi(4415)$:
\begin{equation}
{\cal M}(b \to s \psi^T \to s \gamma )
=i\frac{G_F}{\sqrt 2} ~V_{cb} V_{cs}^* ~a_2 ~\kappa
\sum_i f_{\psi_i}^2 (m_{\psi_i}^2)~\frac{e Q_c}
{m_b}~\bar s ~\sigma_{\mu \nu}~
(1+\gamma_5) ~b ~\epsilon^\mu ~k^\nu\;.\label{eq:mat3}
\end{equation}
The various decay couplings $f_{\psi_i}=f_{\psi_i}(m_{\psi_i}^2) $
are calculated using
\begin{equation}
f_{\psi_i}^2 = \Gamma(\psi_i \to e^+ e^-) \frac{3 m_{\psi_i}}
{Q_c^2 4 \pi \alpha^2},
\end{equation}
which are given in Table I. To extrapolate the coupling
$f_{\psi_i}(k_1^2 =m_{\psi}^2) $ to $f_{\psi_i}(0)$, we use
the suppression factor \cite{ref3}
\begin{equation}
\kappa=\frac{f_{\psi(1S)}^2(0)}{f_{\psi(1S)}^2(m_{\psi}^2)}
=0.12\;,
\end{equation}
obtained from the data on the photoproduction of the $\psi $. This is
taken to be universal for all resonances.
Now, we use Eq.~(\ref{eq:mat3}) to find the matrix element for
$\Lambda_b \to \Lambda \gamma $ through the $b \to s \psi^T \to s 
\gamma $ transition at the quark level, which is given as
\begin{equation}
\langle \Lambda ~\gamma |{\cal H}_{\rm eff}^{\bar c c}|\Lambda_b \rangle
=i\frac{G_f}{\sqrt 2} V_{cb} V_{cs}^* a_2 e Q_c\kappa \epsilon^\mu
k^\nu \sum_i\frac{f_{\psi_i}^2(m_{\psi_i}^2)}{m_b}
\langle \Lambda | \bar s \sigma_{\mu \nu} (1+\gamma_5) b | \Lambda_b
\rangle\;.
\label{eq:eqn13}
\end{equation}

\begin{table}[t]
\begin{center}
\caption{ Values of the vector meson decay constants}
\begin{tabular}{l|cccccc}
\hline\hline
$\psi_i$ & $\psi$(1S) & $\psi$(2S) & $\psi$(3770) 
& $\psi$(4040) & $\psi$(4160) & $\psi$(4415) \\ 
\hline
$f_{\psi_i}$ (GeV) & 0.405 & 0.282 & 0.099 & 0.175 & 0.180 & 0.145\\
\hline
\end{tabular}
\end{center}
\end{table}

\subsection{ Evaluation of the hadronic form factors}

In this section we evaluate the hadronic matrix
elements present in the expression for the decay amplitudes
Eqs.~(\ref{eq:eqn12}) and (\ref{eq:eqn13}).
However, these hadronic matrix elements
involve the tensor currents, and there seems to be 
no known method to evaluate them. 
%
%
%
On the other hand, these elements are easily evaluated\cite{baryon,ref11}
 in COQM 
by taking the overlapping of the initial and final wave functions. 
For $\Lambda_Q$-type baryons, $udQ$,  the  Bargmann-Wigner spinor
function of the third constituent quark $Q$ changes into the spinor
wave functions of the total baryon wave functions.
Thus, the tensor current is calculated as\cite{ref11} 
\begin{equation}
\langle \Lambda |\bar s i \sigma_{\mu\nu}(1 \pm \gamma_5) b |\Lambda_b 
\rangle=
 I_{ud}^{sb}(w) \bar u_f\frac{w+1}{2}i \sigma_{\mu\nu} (1 \pm 
\gamma_5) u_i\;.\label{eq26}
\end{equation}

The single form factor function $I_{ud}^{sb}(w = -v \cdot 
v^\prime)  $ denotes the overlapping of the initial and final
space-time wave function. 
It describes the confinement effects of the quarks 
and is given by
\begin{equation}
I_{ud}^{sb}(w)=\frac{1}{w}~\frac{4 \beta_{\lambda}
\beta_{\lambda}^\prime}
{\beta_{\lambda}+\beta_{\lambda}^\prime }
\frac{1}{\sqrt{C(w)}} \exp(-G(w))\;,
\label{eqnI}
\end{equation}
where
\begin{equation}
C(w)=(\beta_\lambda-\beta_\lambda^{\prime})^2+
4\beta_\lambda \beta_\lambda^{\prime}~w^2\;,
\end{equation}
and
\begin{equation}
G(w)= \frac{4 m_q^2 (\beta_\lambda+\beta_\lambda^\prime)~w (w-1)}
{(\beta_\lambda - \beta_\lambda^\prime )^2 + 4
\beta_\lambda \beta_\lambda^\prime~w^2}\;.
\end{equation}
Here  $\beta_\lambda $ and $\beta_\lambda^\prime $ are
the oscillator strength of the initial and final baryon
oscillator wave functions. They are given in terms of the quark 
masses as
\begin{equation}
\beta_\lambda = \sqrt{\frac{m_q m_b K}{2m_q+m_b}}~~~~~~~~~\mbox{and}
~~~~~~~~~~\beta_{\lambda}^{\prime} =\sqrt{\frac{m_q m_s K}
{2m_q+m_s}}\;.\label{eq:eqn3}
\end{equation}
In the above equation, $m_q$, $m_s $ and $m_b$ denote the masses
of the $u/d$, $s$ and $b$ quarks, and 
$K$ is the universal spring constant for all
hadronic systems with the value $K =0.106$ 
$\mbox{GeV}^3$.\cite{refmR1}

\section{Results and conclusion}

Thus with Eqs.~(\ref{eq:as1}), (\ref{eq:eqn12}), (\ref{eq:eqn13}) 
and (\ref{eq26}) we find the parity conserving and parity
violating ampliudes $a$ and $b$ as
\begin{equation}
a=\frac{G_F}{\sqrt 2} e V_{cb} V_{cs}^* \frac{w+1}{2} I_{ud}^{sb}(w)
\left [ \frac{F_2}{8 \pi^2} (m_b+m_s)+\frac{2}{3} a_2 \kappa \sum_i
\frac{f_{\psi_i}^2(m_{\psi}^2)}{m_b} \right ]
\end{equation}
and
\begin{equation}
b=\frac{G_F}{\sqrt 2} e V_{cb} V_{cs}^* \frac{w+1}{2} I_{ud}^{sb}(w)
\left [ \frac{F_2}{8 \pi^2} (m_b-m_s)+\frac{2}{3} a_2 \kappa \sum_i
\frac{f_{\psi_i}^2(m_{\psi}^2)}{m_b} \right ]
\end{equation}

To estimate the numerical result we have used the following values.
The quark masses used are $m_q=0.4$ GeV, $m_s=0.51$ GeV and $m_b=5$
GeV, which were determined through the analysis of mass 
spectra.\cite{refmR1} 
The particle masses and the lifetime of the $\Lambda_b $ baryon are
taken from Ref.~\citen{ref1}. 
The values of the CKM matrix elements used are
$V_{cb}=0.0395 $ and $V_{cs}=1.04 $. With these values, we obtain the
branching ratio for the decay $\Lambda_b \to \Lambda \gamma $ to be
\begin{equation}
Br(\Lambda_b \to \Lambda \gamma)=0.23 \times 10^{-5}\;.
\label{eq33}
\end{equation}
The main contribution comes from the SD amplitude. By taking only the SD 
amplitude into account, the branching ratio is predicted to be  
$0.215\times 10^{-5}.$\footnote{The branching ratio due only to 
the LD contribution is $1.935\times 10^{-9}$, which is very small
in comparison with the value due only to the SD contribution.}
In obtaining the above branching ratio, Eq.~(\ref{eq33}), 
the value of the form factor function was 
$I_{ud}^{sb}(\omega =2.64)=0.0395$. This extremely small value 
compared with $I=1$ (corresponding to the ``free-quark decay'') 
indicates that the quark-confined effects largely
reduce the decay rate of $\Lambda_b\to\Lambda\ \gamma$.   

The asymmetry parameter is predicted as 
\begin{equation}
\alpha=0.98\;.
\end{equation}

Our predicted branching ratio lies slightly below the 
theoretical value\cite{singer1} $(1 \pm 0.5) \times 10^{-5} $,
while our predicted value of $\alpha$ is consistent with 
the theoretical value\cite{singer1} 0.9.
Additional experimental data would greatly help for a better
understanding of the weak radiative decays of heavy baryons, which
can serve as a signal of new physics beyond the Standard Model.

This decay process was previously studied by
Cheng et al.\cite{cheng2} 
They took into account the short distance contribution only
and estimated the branching ratio following two different approaches.
In the first method they treated both the $b$ and $s$ quarks as heavy
and included a correction of order $1/m_s $. However, the $1/m_s$
correction to the $\Lambda_b \to \Lambda \gamma $ amplitude is about
$50 \% $ for $m_s=510 $ MeV which is quite sizable. Hence it is
important to include higher order $1/m_s $ corrections. In the
second method they treated the $b$ quark as heavy and
used the MIT bag model to calculate the form factors at zero recoil 
and then extrapolate them to $q^2=0 $ by assuming a dipole $q^2$
dependence of the form factor. In this paper we have
considered both the short and long distance contributions and
taken into account the confined effects by using the COQM, 
which has proven to be very successful for
the phenomenology of both the heavy and light hadrons. It should be 
noted that in this model we do not need to extrapolate the form factor
to the particular point of interest, as the model gives the full
spectrum for the hadronic form factor functions. Thus our model provides a
better theoretical understanding of the rare decays of heavy baryons.

\acknowledgements

R. M. would like to thank CSIR, Govt. of India, for a fellowship.
A. K. G. and M. P. K. acknowledge financial support from
DST, Govt. of India.

\end{document}